\documentclass[aps,twocolumn]{revtex4}

\usepackage{amsmath,amssymb,graphicx}
\usepackage{algorithmic}
\usepackage{enumerate}
\usepackage{color}

\newcommand{\g}{j}

\newcommand{\wn}{\tau}
\newcommand{\F}{{\tilde F}}

\newcommand{\wt}{Q}

\newcommand{\bc}{\zeta}
\newcommand{\tbc}{TBC}
\newcommand{\f}{f}
\newcommand{\lam}{\lambda}
\newcommand{\cJ}{\cal J}
\newcommand{\po}{I}
\newcommand{\poc}{I_{\cal J}}

\begin{document}

\title{Evidence against a mean-field description of short-range spin
glasses revealed through thermal boundary conditions}

\author{Wenlong Wang}
\email{wenlong@physics.umass.edu}
\affiliation{Department of Physics, University of Massachusetts, Amherst, Massachusetts 01003, USA}

\author{Jonathan Machta}
\email{machta@physics.umass.edu}
\affiliation{Department of Physics, University of Massachusetts, Amherst, Massachusetts 01003, USA}
\affiliation{Santa Fe Institute, 1399 Hyde Park Road, Santa Fe, New Mexico 87501, USA}

\author{Helmut G. Katzgraber}
\affiliation{Department of Physics and Astronomy, Texas A\&M University, College Station, Texas 77843-4242, USA}
\affiliation{Materials Science and Engineering Program, Texas A\&M University, College Station, Texas 77843, USA}
\affiliation{Santa Fe Institute, 1399 Hyde Park Road, Santa Fe, New Mexico 87501, USA}

\begin{abstract}

A theoretical description of the low-temperature phase of short-range
spin glasses has remained elusive for decades. In particular, it is
unclear if theories that assert a single pair of pure states, or
theories that are based on infinitely many pure states---such as
replica symmetry breaking---best describe realistic short-range systems.
To resolve this controversy, the three-dimensional Edwards-Anderson
Ising spin glass in {\em thermal} boundary conditions is studied
numerically using population annealing Monte Carlo.  In thermal boundary
conditions all eight combinations of periodic vs antiperiodic boundary
conditions in the three spatial directions appear in the ensemble with
their respective Boltzmann weights, thus minimizing finite-size
corrections due to domain walls.  From the relative weighting of the
eight boundary conditions for each disorder instance a {\em sample
stiffness} is defined, and its typical value is shown to grow with
system size according to a stiffness exponent. An extrapolation to the
large-system-size limit is in agreement with a description that supports
the droplet picture and other theories that assert a single pair of pure
states. The results are, however, incompatible with the mean-field
replica symmetry breaking picture, thus highlighting the need to go
beyond mean-field descriptions to accurately describe short-range
spin-glass systems.

\end{abstract}

\maketitle

\section{Introduction}

A plethora of problems across disciplines and, in particular, a wide
variety of optimization problems map onto spin-glass-like Hamiltonians
\cite{binder:86,mezard:87,young:98,hartmann:04,stein:13,lucas:14}.  As such,
despite the fact that only a selected class of disordered magnets, such
as ${\rm LiHoF_4}$ or ${\rm Au_x Fe_{1-x}}$ show this intriguing state
of matter, spin glasses have been of great importance across multiple fields including condensed matter
physics, evolutionary biology, neuroscience and computer science. Most recently spin glasses have played a
pivotal role in the development of new computing prototypes based on
quantum bits in both a theoretical, as well as device-centered role. For
example, the stability of topologically protected quantum computing
proposals \cite{freedman:02,kitaev:03,bombin:06} against different error
sources---recently implemented experimentally \cite{nigg:14a}---heavily
relies on spin-glass physics \cite{dennis:02,katzgraber:09c,bombin:12}.
Similarly, the native benchmark problem currently used to gain a deeper
understanding of state-of-the-art quantum annealing machines is based on
a spin-glass Hamiltonian
\cite{dickson:13,boixo:13a,katzgraber:14,ronnow:14a}. Given this recent
renaissance of spin glasses to benchmark novel algorithms, as well as to
develop cutting-edge computing paradigms, it is unsettling that no
consensus exists as to whether mean-field theory, also known as replica
symmetry breaking theory (RSB)
\cite{parisi:79,parisi:80,parisi:83,rammal:86,mezard:87,young:98},
accurately describes the low-temperature phase of these systems.

Spin-glass models are well understood in the mean-field regime where
infinite-range interactions dominate. However, in finite space
dimensions spin glasses are still poorly understood and have been the
subject of a long-standing controversy. In this paper we seek to help resolve
this controversy. Using numerical methods, we study the low-temperature
phase of the three-dimensional Edwards-Anderson (EA) spin glass
\cite{edwards:75}. In so doing, we introduce two methods that promise to
be useful in the study of other disordered systems---thermal boundary
conditions and sample stiffness extrapolation.

The controversy concerning the EA model is between two competing classes
of theories as to the nature of the low-temperature phase.  One
proposal, championed by Parisi and collaborators
\cite{parisi:79,parisi:80,parisi:83,rammal:86,mezard:87,young:98,parisi:08},
is that finite-dimensional EA spin glasses behave like the mean-field
Ising spin glass, known as the Sherrington-Kirkpatrick (SK) model
\cite{sherrington:75}.  Parisi analytically studied the SK model
\cite{parisi:79,parisi:80,parisi:83} and found that the low-temperature
phase is characterized by an unusual form of symmetry breaking called
replica symmetry breaking (RSB). This RSB solution of the SK model
predicts that there is an infinity of pure thermodynamic states and that
the overlap distribution of these pure states is not self-averaging.
Many features of Parisi's RSB solution of the SK model have now been
verified by rigorous mathematical methods \cite{Pan12}.  The mean-field or RSB picture
for finite-dimensional EA spin glasses asserts that the qualitative
features of the SK model also hold for finite-dimensional models so
that, in particular, there are infinitely many pure states in the
thermodynamic limit.

In contrast to the RSB picture, the main competing class of theories for the
three-dimensional (3D) EA model assume that the low-temperature phase
consists simply of a single pair of pure states related by the
spin-reversal symmetry of the Hamiltonian.  The earliest and most widely
accepted of these theories is the 
``droplet picture'' developed by McMillan \cite{mcmillan:84b}, Bray and
Moore \cite{bray:86}, and Fisher and Huse
\cite{fisher:86,fisher:87,fisher:88}. The droplet picture asserts that
the low lying excitations of the pure states are compact droplets with
energies that scale as a power of the size of the droplet.  By contrast,
the low lying excitations in the RSB picture are space filling objects.

Several features of the original RSB picture for finite-dimensional EA
models have been mathematically ruled out in a series of papers by
Newman and Stein \cite{newman:92,newman:96,NeSt98,NeSt02,newman:03}.
These authors provide two alternative theories for finite-dimensional EA
models, both of which have infinitely many pure states. The first is a
nonstandard RSB picture, similar to the original RSB picture but with a
self-averaging thermodynamic limit.  Newman and Stein give heuristic
arguments against the nonstandard RSB picture but do not rule it out.   On the other hand, the nonstandard RSB picture is promoted as a viable theory for finite-dimensional EA models in Ref.\ \cite{read:14a}.
The second is the ``chaotic pairs'' picture. Here there are infinitely
many pure states but they are organized in such a way that in each
finite volume only a single pair of states related by a global spin flip
is seen.

In the following we refer to all pictures that display a single pair of
pure states in each large finite volume as {\em two-state} pictures.
Therefore, the droplet and chaotic pairs pictures are both two-state
pictures within this definition. Note that for the droplet model it is
the same pair of states in every volume while for chaotic pairs a
different pair of states is manifest in each volume.

Parallel to these analytical efforts, many computational studies have
been aimed at distinguishing between the two classes of theories 
(see,
for  example,
Refs.~\cite{palassini:99,katzgraber:01,katzgraber:02,alvarez:10a,houdayer:00,krzakala:00,katzgraber:03,yucesoy:12,billoire:14a}).
Unfortunately, computational methods have  been difficult to apply to
spin glasses. The fundamental questions concern the limit of large
system sizes, however, attempts to extrapolate to large sizes have not been
conclusive because the range of sizes accessible to simulations {\em at low
temperatures} is quite small and, for fixed size, the variance between
samples for many observables is quite large. Thus, a straightforward
extrapolation to large sizes based on mean values of observables can be
misleading. Computational studies have yielded a confusing
mixture of results: Some point to the RSB picture, some to a two-state
picture, and some to a mixed scenario, known as the the  ``trivial
nontrivial'' (TNT)
picture described in Refs.~\cite{palassini:00,krzakala:00,katzgraber:01}.
Recently, there have been efforts to analyze statistics other than
simple disorder averages
\cite{yucesoy:12,billoire:13,YuKaMa13b,Middleton13,MoGa13,BiMa14} but these
methods have not been definitive either and so the controversy
continues.

Here we introduce two related innovations to more effectively
extrapolate from small system sizes to the large system-size limit.
First, we employ {\em thermal} boundary conditions instead of the usual
periodic boundary conditions. In $d$ space dimensions, thermal boundary
conditions allow the $2^d$ combinations of periodic or antiperiodic
boundary conditions each to appear with the correct Boltzmann weights.
The idea is to let the system choose boundary conditions that minimize
the presence of domain walls and thus finite-size effects. Second, based
on thermal boundary conditions, we define a spin stiffness measure for
each sample.  We show, as expected for a low-temperature phase, that the
sample stiffness becomes large as the system size becomes large.  We
then study the behavior of the system in the limit of large sample
stiffness and relate the system's behavior for large sample stiffness to
the large system-size limit. During this process we also obtain new
measurements of the spin stiffness exponent at nonzero temperatures.
These generic techniques promise to be of broad utility in
understanding disordered systems in statistical mechanics, not just the
EA spin-glass model. Furthermore, using this approach we conclude
that a two-state picture best describes the low-temperature phase of the
three-dimensional Edwards-Anderson Ising spin glass.

The paper is structured as follows. In Sec.~\ref{sec:model} we
introduce the studied model, as well as thermal boundary conditions.
Section \ref{sec:methods} describes the implementation of population
annealing Monte Carlo used in this study, followed by the measured
quantities in Sec.~\ref{sec:obs}. Results are presented in
Sec.~\ref{sec:results}, followed by a discussion in Sec.~\ref{sec:disc}
and concluding remarks.

\section{The Edwards-Anderson model in Thermal Boundary Conditions}
\label{sec:model}

\subsection{Edwards-Anderson model}

We study the three-dimensional Edwards-Anderson Ising spin-glass model
\cite{edwards:75}.  The model is defined by the Hamiltonian
\begin{equation}
{\mathcal H} = - \sum_{\langle i,j \rangle} J_{ij} S_i S_j ,
\label{eq:ham}
\end{equation}
where $S_i \in \{\pm 1\}$ are Ising spins and the sum is over nearest
neighbors on a cubic lattice of linear size $L$. The random couplings
$J_{ij}$ are chosen from a Gaussian distribution with zero mean and unit
variance.  A set of couplings ${\cJ} = \{ J_{ij} \}$ defines a disorder
realization or ``sample.''

\subsection{Thermal boundary conditions}
\label{sec:tbc}

Most Monte Carlo simulations of spin systems are performed with periodic
boundary conditions (PBC) because it is often assumed that periodic
boundary conditions yield the mildest finite-size correction. Free
boundary conditions are sometimes also employed \cite{katzgraber:02} as
are antiperiodic boundary conditions in one direction for the purpose of
measuring spin stiffness. In this work we consider {\it thermal boundary
conditions} (\tbc).  Thermal boundary conditions include the set of all
$2^d$ choices of periodic or antiperiodic boundary conditions in $d$
spatial dimensions. This means that for three space dimensions ($d = 3$)
we have 8 possible choices.  For example, one of the 8 elements in
the set of boundary conditions is ``periodic in the $x$-direction and
antiperiodic in the $y$ and $z$ directions.''  For each boundary
condition $\bc$, there is a free energy $F_\bc$ and the probability
distribution for spin states in \tbc~is the weighted mixture of the
eight boundary conditions with weights $e^{-\beta F_\bc}$.  An
equivalent way to describe thermal boundary conditions is to say that
the eight boundary conditions are annealed so that each spin
configuration together with a boundary condition appears with its proper
Boltzmann weight.

The motivation for using thermal boundary conditions for spin glasses
can be explained by considering two simpler examples---the ferromagnetic
and antiferromagnetic Ising models on a square lattice with lattice size
$L$ an odd number. For the ferromagnetic Ising model  periodic boundary
conditions in all directions are natural and appropriate because they do
not induce domain walls in the ordered phase.  However, for the
antiferromagnetic Ising model, periodic boundary conditions will induce
domain walls and the observables for finite systems will have strong
finite-size corrections. The natural boundary conditions for the
antiferromagnet with $L$ odd are antiperiodic in all directions.  Now,
suppose we are asked to simulate an Ising model but we are not told
whether it is a ferromagnet or antiferromagnet. If we use thermal
boundary conditions then we will automatically choose the natural
boundary conditions independent of which model we have been given,
namely periodic in all directions if the system is a ferromagnet and
antiperiodic in all directions if the system is an antiferromagnet.  The
other boundary conditions will induce domain walls and therefore have
higher free energies. The difference in free energy between thermal
boundary conditions and any of the domain-wall-inducing boundary
conditions scales as $L^\theta$ where $\theta$ is the spin stiffness
exponent. For the Ising model (either ferromagnetic or
antiferromagnetic) in the low-temperature phase, $\theta = d - 1 \ge 0$,
and, even for modest system sizes, thermal boundary conditions are
essentially the same as the single natural boundary condition because
all unfavorable choices are suppressed.

While one can {\em a priori} determine the optimal boundary conditions for
simple systems such as ferromagnets and antiferromagnets, the same is not true for spin glasses.
For a given sample, a single boundary condition such as PBC may induce
domain walls and induce large finite-size effects. The motivation for
using thermal boundary conditions is thus the same as for the simple
(anti)ferromagnetic example discussed above. Because we do not know
which of the eight periodic/antiperiodic boundary conditions fits the
sample best, we simply let the system choose by minimizing the free
energy.

At zero temperature, thermal boundary conditions correspond to selecting
from among the $2^d$ boundary conditions those with the lowest energy
ground states. These boundary conditions have been employed with exact
algorithms for finding ground states of two-dimensional spin glasses
\cite{LaCo02,thomas:07}.  Thomas and Middleton \cite{thomas:07} call
thermal boundary conditions ``extended'' boundary conditions and argue,
as we do, that these boundary conditions minimize finite-size effects. Similar ideas but using periodic and antiperiodic boundary conditions in a single direction are discussed in \cite{hukushima:99,sasaki:05,sasaki:07b,HasenbuschTBC}.

For the mathematical statistical physicist the difficulties produced by
spurious domain walls are avoided by using ``windows:''  Consider a very
large system of linear size $L$ and a large window inside this system of
linear size $\ell$ such that $1 \ll \ell \ll L$. Then any domain walls
induced by the ``bad'' boundary conditions will almost surely lie
outside the window and observables measured within the window will not
be influenced by the domain wall. By collecting data only inside the
window the results are then independent of the boundary conditions.
Unfortunately, the computational statistical physicist does not have the
luxury of collecting equilibrated data in this way for spin glasses
where attainable system sizes deep within the low-temperature phase do
not exceed, for example, $L \approx 10$ in three space dimensions.

Thermal boundary conditions may also be used to measure the spin
stiffness exponent $\theta$ by comparing the free energy of \tbc~with
other boundary conditions. For example, for spin glasses it is
sufficient to compare the free energy for \tbc~with that for PBC. This
approach is expected to yield the same exponent but a different
prefactor for the spin stiffness as compared to the standard method of
taking the absolute value of the free energy difference between periodic
and antiperiodic boundary conditions.

\section{Methods}
\label{sec:methods}

We use population annealing Monte Carlo \cite{HuIb03,Mac10a,MaEl11} to
simulate the EA model. The most common method for large-scale spin-glass
simulations is parallel tempering Monte Carlo \cite{hukushima:96},
however, parallel tempering Monte Carlo is not well suited to thermal
boundary conditions and is not easily used to measure free energies.  In
contrast, population annealing is able to simulate \tbc~ and accurately
measure free energies in a straightforward way.

Population annealing is related to simulated annealing where the
temperature of the system is lowered in a step-wise fashion following a
predefined annealing schedule. In each annealing step a Markov chain
Monte Carlo algorithm is applied to the system at the current
temperature in the schedule. In population annealing a large population
of replicas of the system are simultaneously annealed from high to low
temperature.  However, simulated annealing is designed  for finding only
ground states and it does not correctly sample equilibrium states at the
temperatures traversed during the annealing schedule.  Population
annealing corrects this deficiency by adding a resampling step that
ensures that the population of replicas at every temperature is a Gibbs
ensemble at that temperature. Population annealing is an example of a
sequential Monte Carlo algorithm \cite{DoFrGo01} and it converges to the
correct equilibrium distribution as the population size increases. It is
thus well suited to parallel computation and our implementation uses
OpenMP.

The algorithm works as follows. Let $R_0$ be the initial size of the
population of replicas of the system.  In our implementation of
population annealing, each replica is initialized independently at
infinite temperature $T$ ($\beta=1/T=0$).  Each replica has the same set
of couplings.  For thermal boundary conditions, $1/8$ of the replicas
are assigned to each of the $8$ boundary conditions.  The temperature of
the population is now cooled from $\beta=0$ in a sequence of steps to a
target temperature $\beta_0$. The annealing step from $\beta$ to
$\beta^\prime$ ($\beta^\prime>\beta$) consists of two stages: The first
stage is resampling and the second stage is the application of the
Metropolis algorithm at inverse temperature $\beta^\prime$. In the
resampling step, some replicas are eliminated and others are duplicated.
For \tbc, when a replica is copied, its boundary condition is copied
with it.

The resampling step works as follows. Suppose we have
$\tilde{R}_\beta$ replicas that represent an equilibrium ensemble at
inverse temperature $\beta$ and we want to lower the temperature to
$\beta^\prime>\beta$. The ratio of the statistical weight at $\beta$ to
$\beta^\prime$ for replica  $\g$, with energy $E_\g$ is $\exp\left[
-(\beta^\prime-\beta)E_\g\right]$. In principle this factor should
represent how many copies to make of the system.  However, this ratio is
typically larger than unity. In order to keep the population size
roughly fixed we need to normalize the ratio.  First compute normalized
weights $\wn_\g(\beta,\beta^\prime)$ whose sum over the ensemble is
$R_0$,
\begin{equation}
\label{eq:wn}
\wn_\g(\beta,\beta^\prime)=
\left(\frac{R_0}{\tilde{R}_\beta}\right)
\frac{\exp\left[-(\beta^\prime-\beta)E_{\g}\right]}{\wt(\beta,\beta^\prime)},
\end{equation}
where $\wt$ is the normalization given by
\begin{equation}
\label{eq:Q}
\wt(\beta,\beta^\prime)=
\frac{\sum_{j=1}^{\tilde{R}_\beta} 
\exp\left[-(\beta^\prime-\beta)E_{\g}\right]}{\tilde{R}_\beta} .
\end{equation}
The new population at temperature $\beta^\prime$ is obtained by
differential reproduction.  The number of copies $n_{\g}$ of replica
$\g$ is either the floor (greatest integer less than)  $\lfloor \wn_\g
\rfloor$ or ceiling $\lceil \wn_\g \rceil$ with probabilities $\lceil
\wn_\g \rceil-\wn_\g$ and $\wn_\g-\lfloor \wn_\g \rfloor$, respectively.
Note that this choice ensures that the expectation of $n_{\g}$ is
$\wn_\g$.  It also minimizes the variance of $n_{\g}$ among all integer
probability distributions with this expectation. (Note that it is possible to have $n_{\g}=0$ for replica $\g$.) In our implementation,
the size of the population at each temperature $\tilde{R}_\beta$ is
variable but stays close to the target value $R_0$.  In the next stage of
the annealing step, every member of the new population is subject to
$N_S$ sweeps of the Metropolis algorithm at inverse temperature
$\beta^\prime$.  Our implementation of population annealing follows a schedule of $N_T$ inverse temperatures $\beta$ that are evenly spaced between $\beta=0$ and $\beta_0$. The system sizes and temperatures in
the simulations, together with  the parameters of the population
annealing simulations are shown in Table \ref{tab:param}. The definition of hard samples is given below. Note that although the number of sweeps per temperature $N_S$ is small, the total number of sweeps, $R_0N_SN_T$, is large and comparable to the number of sweeps performed in parallel tempering simulations. In population annealing, equilibration results from large values of $R_0$ and is guaranteed in the limit $R_0 \rightarrow \infty$ for fixed $N_S$ and $N_T$.

The free energy can be estimated \cite{Mac10a} from the normalization factors
$\wt(\beta,\beta^\prime)$ defined in Eq.~\eqref{eq:wn} according to
\begin{equation}
\label{eq:sumQ}
-\beta_k \F(\beta_k) = 
\sum_{\ell=N_T-1}^{k+1} 
\ln \wt(\beta_{\ell},\beta_{\ell-1}) + \ln \Omega ,
\end{equation}
where $\Omega$ is the number of configurations of the system so that for
$N$ Ising spins, $\Omega=2^N$ for PBC and $\Omega=2^{(N+d)}$ for \tbc~in
$d$ dimensions.

\begin{table}
\caption{
Parameters of the numerical simulations for different system sizes $L$
and periodic (PBC), as well as thermal (TBC) boundary conditions. $R_0$
represents the number of replicas, $1/\beta_0$ is the lowest temperature
simulated, $N_T$ the number of temperatures used in the annealing
schedule, $N_S$ the number of sweeps per temperature, and $M$ the number
of samples. $M_{\rm PBC}$ is the number of hard samples for periodic
boundary conditions and $M_{\rm TBC}$ the number of hard samples for
thermal boundary conditions.
\label{tab:param}
}
\begin{tabular*}{\columnwidth}{@{\extracolsep{\fill}} l c c c c c c l l}
\hline
\hline
$L$  & ${R_0}_{PBC}$ & ${R_0}_{TBC}$ & $1/\beta_0$ & $N_T$ & $N_S$ & $M$ & $M_{\rm PBC}$ & $M_{\rm TBC}$\\
\hline
$4$  & $5\,10^4$ & $5\,10^4$ & $0.200$     & $101$ & $10$  & $4941$ & $0$	&   $0$ \\
$6$  & $2\,10^5$ & $2\,10^5$ & $0.200$     & $101$ & $10$  & $4959$ & $0$   &   $0$ \\
$8$  & $5\,10^5$ & $5\,10^5$ & $0.200$     & $201$ & $10$  & $5099$ & $5$   &  $33$ \\
$10$ & $   10^6$ &$   2\,10^6$ & $0.200$     & $301$ & $10$  & $4945$ & $286$ & $291$ \\
$12$ & $   10^6$ &$   3\,10^6$ & $0.333$     & $301$ & $10$  & $5000$ & $533$ & $386$ \\
\hline
\hline
\end{tabular*}
\end{table}

Because population annealing has not been used before for large-scale
simulations in statistical physics, we did a careful comparison  to data
previously obtained using parallel tempering \cite{yucesoy:12,YuKaMa13}.
We measured observables for the same set of samples studied in
Refs.~\cite{yucesoy:12,YuKaMa13}, comprising approximately $5000$
samples for each system size $L$ with periodic boundary conditions. We
found no statistical difference between population annealing and
parallel tempering for any disorder averaged observable (see Sec.\
\ref{sec:I}  for a comparison of one observable).  A detailed
 comparison between population annealing and parallel tempering will be
 presented in a subsequent publication
\cite{WaMaKaXX}.

The convergence to equilibrium of population annealing for each sample
can be quantified using the {\em family entropy}.  Define {\em family}
$i$ as the set of replicas at some low temperature that are descended
from replica $i$ at the highest temperature.  In practice most families
are empty sets.  Let $\eta_i$ be the fraction of the 
population in family $i$, i.e., the fraction of the population at the
low temperature that is descended from replica $i$ in the initial, high-temperature
population.  Then the family entropy $S_f$ is given by
\begin{equation}
S_f = -\sum_i \eta_i \log \eta_i .
\end{equation}
The exponential of $S_f$ is an effective number of families.  For
example, if there are $k$ surviving families all of the same size then
$\eta_i = 1/k$ for each surviving family and $e^{S_f}=k$.  In practice,
the family sizes are exponentially distributed.

Since each family has an independent history during the simulation,
$e^{-S_f/2}$ is a conservative measure of statistical errors.  As
discussed in Ref.~\cite{WaMaKaXX}, $e^{-S_f}$ is a reasonable measure of
systematic errors. In our simulations, we require that $e^{-S_f}<0.01$
for every disorder sample. For hard samples that do not meet this
requirement with the standard population size given in Table
\ref{tab:param} we increased the population size until this
equilibration criterion was met. The numbers of hard samples $M_{PBC}$ and $M_{TBC}$ for periodic and thermal boundary conditions, respectively, are given in
Table \ref{tab:param}.  Most hard samples were equilibrated
using $5$ runs with $R_0=3\times10^6$, which were then combined using weighted
averaging \cite{Mac10a} for a total population of $1.5\times10^7$. For the
hardest samples of size $L = 10$ and $12$, population sizes up to $10^8$
were required to meet the equilibration criterion.

\section{Measured Quantities}
\label{sec:obs}

\subsection{Free energy, ground-state energy and spin stiffness}

Using Eq.~\eqref{eq:sumQ} we measure the free energies, $F_{\cJ}^{\rm
\tbc}$ and $F_{\cJ}^{\rm PBC}$, for each sample $\cJ$ in thermal (\tbc)
and periodic  (PBC) boundary conditions, respectively.  We also measure
the ground-state energies $E^{\rm \tbc}_{\cJ}$ and $E^{\rm PBC}_{\cJ}$
for each sample in both \tbc~and PBC, respectively. We compute the
ground-state energy by taking the minimum energy in the population at
the lowest temperature ($T=0.2 \ll T_c$). We report on a careful study
of the ground-state calculation in a separate paper \cite{WaMakaXY}.
There we show that the average ground-state energy agrees with other
methods, that multiple runs always yield the same ground state, and that
for a small number of the hardest samples we find agreement with exact
branch and bound methods.

The traditional measure of spin stiffness is the difference between the
free energy, or at zero temperature, the ground-state energy of two
different boundary conditions---usually periodic and antiperiodic in a
single direction with periodic boundary conditions in all other
directions. For spin glasses, this quantity may be of either sign and
the absolute value must be taken before performing the disorder average.
Here we consider the free energy (ground-state energy) difference
between thermal boundary conditions and periodic boundary conditions.
This quantity is nonnegative because periodic boundary conditions are
contained in the \tbc~ ensemble of boundary conditions so no absolute
value needs to be taken. We refer to $\Delta F$ as the disorder average
free energy (ground-state energy) difference between \tbc~ and PBC.  The
scaling of $\Delta F$ with system size $L$ defines the stiffness
exponent $\theta$, 
\begin{equation} 
\Delta F \sim L^\theta .
\end{equation} 
We measure $\theta$ at $T=0$, $0.2$, and $0.42$ by fitting to this equation.

The free energy for each boundary condition in the \tbc~ensemble can
be measured using an analog of Eq.~\eqref{eq:sumQ} by partitioning $\wt$
into its 8 boundary condition components but we did not collect data
to do this measurement. Instead, we estimate the ratio of the free
energy of the dominant boundary condition to the free energy of all the
other boundary conditions combined. Let $\f_{\cJ}$ be the fraction of
the population in the boundary condition with the largest population in sample
$\cJ$.  The quantity $\lam_{\cJ}$,
\begin{equation}
\label{eq:lam}
\lam_{\cJ} = \log\frac{f_{\cJ}}{(1-f_{\cJ})} ,
\end{equation} 
is an estimator of the free-energy difference (times $-\beta$) between
the dominant boundary condition and all other boundary conditions
in sample $\cJ$. Note that $\lam_{\cJ}$ is a measure of the stiffness of
sample $\cJ$.  If only one boundary condition dominates the ensemble of
boundary conditions it means that inserting a domain wall is very costly
and the sample is stiff while if $\lam_{\cJ}$ is small, the domain walls
induced by changing boundary conditions have little cost and the sample
is not stiff.  Note that, in principle, all boundary conditions could have
equal weight so $\lam \geq -\log 7$. 

\subsection{Order parameter distribution}

The order parameter for spin glasses is obtained from the spin overlap,
$q$ defined by
\begin{equation}
\label{eq:q}
q= \frac{1}{N} \sum_i S_i^{(1)}S_i^{(2)},
\end{equation}
where the superscripts ``(1)'' and ``(2)'' indicate two statistically
independent spin configurations chosen from the Gibbs distribution. Let
$P_{\cJ}(q)$ be the overlap distribution for sample $\cJ$ and let $P(q)$
be the disorder average of the overlap distribution. In population
annealing, the pairs of independent spin configurations used in
Eq.~\eqref{eq:q} are chosen randomly from the population of replicas
with the restriction that the two replicas are from {\em different}
families. This ensures that the spin configurations are independent.

Two-state pictures make very different predictions from the RSB picture
for $P_{\cJ}(q)$ for the low-temperature phase of the EA model in the
infinite-volume limit. If there is a single pair of pure states then
$P_{\cJ}(q)$ consists of two $\delta$ functions at $\pm q_{\rm EA}$ and,
of course, $P(q)$ after disorder averaging is the same.  Here $q_{\rm
EA}$ is the Edwards-Anderson order parameter. In the RSB picture,
$P_{\cJ}(q)$ consists of a countable infinity of $\delta$ functions of
varying weights densely filling the range between $\pm q_{\rm EA}$ while
$P(q)$ is a smooth function between $\pm q_{\rm EA}$ with
delta functions at $\pm q_{\rm EA}$.  Thus, one can, in principle,
distinguish between the two classes of pictures by examining $P(q)$ near
the origin (and thus away from $q_{\rm EA}$).  A measure of the weight
near the origin of the overlap distribution is $\po_{\cJ}(q)$,
\begin{equation}
\po_{\cJ}(q) = \int_{-q_0}^{+q_0}  dq P_{\cJ}(q) .
\end{equation}
We refer to the disorder average of $\po_{\cJ}(q)$ as $\po(q)$.  The
choice $q_0=0.2$ has been used in many past studies
\cite{katzgraber:01,yucesoy:12} to distinguish the RSB and two-state
pictures and, in this work, we investigate the statistics of
$\po_{\cJ}(q_0 = 0.2)$. In the following, we use the symbols $\poc$ and
$\po_L$ as abbreviations for  $\po_{\cJ}(0.2)$ and its disorder average
$\po(0.2)$ for size $L$, respectively.

\section{Results}
\label{sec:results}

In this section, we present results for the spin stiffness
(\ref{sec:Ia}), the order parameter distribution near zero, $I_{\cJ}$
(\ref{sec:I}) and the correlation of $I_{\cJ}$ and $\lambda_{\cJ}$
(\ref{sec:Ic}). The main result of this section is that $\lambda_{\cJ}$
increases with system size and that stiff samples have small values of
$I_{\cJ}$.

\subsection{Spin stiffness}
\label{sec:Ia}
\begin{figure}[htb]
\begin{center}
\includegraphics[scale=0.68]{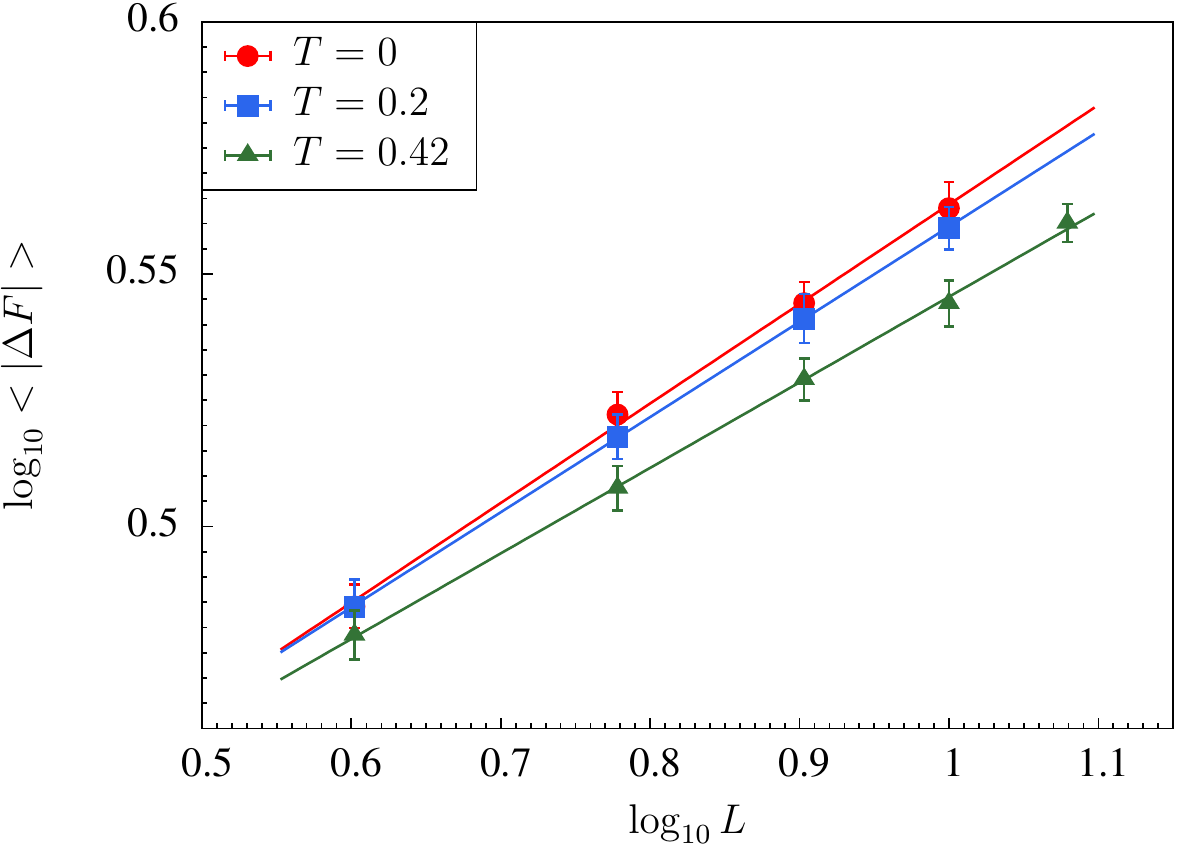}
\caption{
Free-energy change $\Delta F$ vs system size $L$ for $T=0$, $0.2$, and
$0.42$.  The straight lines are fits of the form $\Delta F \sim a
L^\theta$.
}
\label{fig:theta}
\end{center}
\end{figure}

Figure \ref{fig:theta} shows the free-energy difference or, for $T=0$,
ground-state energy difference $\Delta F$ between \tbc~and PBC for
temperatures $T=0$, $0.2$, and $0.42$ as a function of system size $L$.
The straight lines are best fits to the functional form $\Delta F \sim a
L^\theta$. The fits for $\theta$ are shown in Table \ref{tab:theta}.  The
result $\theta(T=0)= 0.197 \pm 0.017$ is in reasonable agreement though
at the low end of previous measurements of $\theta$ carried out at zero
temperature \cite{Hartmann99,palassini:99,CaBrMo02,Boettcher04}. Note that the
stiffness exponent has not previously been measured at nonzero
temperature. We see that $\theta$ decreases as temperature increases.
Presumably, this is a finite-size effect because $\theta$ is expected to
have a single asymptotic value throughout the low-temperature
phase~\cite{fisher:88}.

\begin{table}
\caption{
Estimates of the stiffness exponents $\theta$ and $\theta_\lam$ for
different temperatures $T$.
\label{tab:theta}
}
\begin{tabular*}{\columnwidth}{@{\extracolsep{\fill}} l l l l}
\hline
\hline
$T$ 		& $0$         & $0.2$       & $0.42$ \\
\hline
$\theta$ 	& $0.197(17)$ & $0.189(17)$ & $0.169(12)$ \\
$\theta_\lam$   & ---         & $0.290(30)$ & $0.268(20)$ \\ 
\hline
\hline
\end{tabular*}
\end{table}

\begin{figure}[htb]
\begin{center}
\includegraphics[scale=0.68]{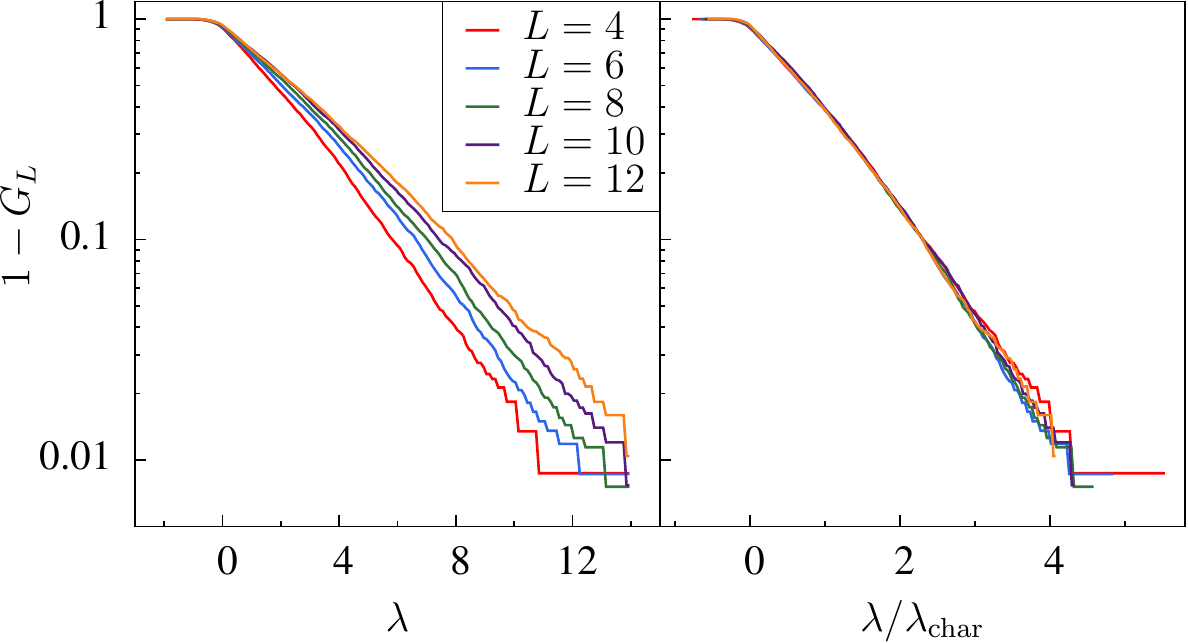}
\caption{
(Left panel) Linear-log plot of $1-G_L(\lam)$ (the complementary
cumulative distribution function) vs $\lam$ for sizes $L=4$ through $12$
at $T=0.42$. (Right panel) $1-G_L(\lam/\lam_{\rm char}(L))$ vs
$\lam/\lam_{\rm char}(L)$.
}
\label{fig:lamccdf42}
\end{center}
\end{figure}

\begin{figure}[htb]
\begin{center}
\includegraphics[scale=0.68]{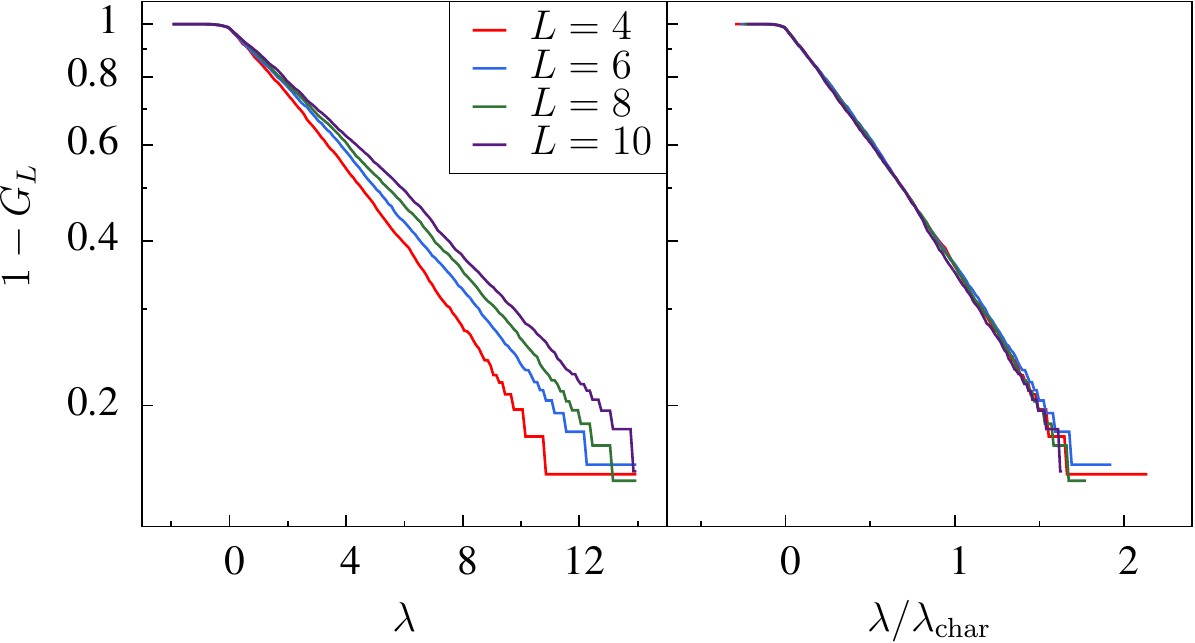}
\caption{
(Left panel) Linear-log plot of $1-G_L(\lam)$ (the complementary
cumulative distribution function) vs  $\lam$ for sizes $L=4$ through
$10$ at $T=0.2$. (Right panel) $1-G_L(\lam/\lam_{\rm char}(L))$ vs
$\lam/\lam_{\rm char}(L)$.
}
\label{fig:lamccdf20}
\end{center}
\end{figure}

Next consider the sample stiffness measure $\lam_{\cJ}$, defined in
Eq.~\eqref{eq:lam}. Let $G_L(\lam)$ be the cumulative distribution
function for $\lam$. The left panel of Fig.~\ref{fig:lamccdf42} is a log
plot of $1-G_L(\lam)$, the complementary cumulative distribution
function, for $\lam$ at $T=0.42$, and sizes $4$ through $12$.  The
nearly straight line behavior of $\log(1-G_L(\lam))$ is indicative of a
nearly exponential tail and suggests a data collapse if $\lam$ is scaled
by a characteristic $\lam_{\rm char}(L)$ given by the slope of the line.
Since the tail is not perfectly straight, we instead define $\lam_{\rm
char}(L)$ in terms of median-like quantities.  If the distribution were
exactly exponential then $1-G(\lam) = e^{-\lam/\lam_{\rm char}}$ and
$1-G(\lam_{\rm char} \log b) = 1/b$ for any $b$.  If the distribution is
not perfectly exponential, $\lam_{\rm char}$ depends on $b$ so there is
some ambiguity in the definition.  We choose $b$ such that $\lam_{\rm
char}$ is obtained from the tail of the distribution but not so far into
the tail that the statistics are poor. For $T=0.2$ we choose $b=2$ so
$\lam_{\rm char}$ is defined as the median divided by $\log 2$.  For
$T=0.42$ we choose $b=10$ to ensure that $\lam_{\rm char}$ is obtained
from the tail of the distribution.  The right panels of
Figs.~\ref{fig:lamccdf42} and~\ref{fig:lamccdf20} show
$1-G_L(\lam/\lam_{\rm char}(L))$ for $T=0.42$ and $T=0.2$, respectively,
and reveal that all of the cumulative distributions collapse onto the
same curve when scaled by $\lam_{\rm char}(L)$.

\begin{figure}[htb]
\includegraphics[scale=0.68]{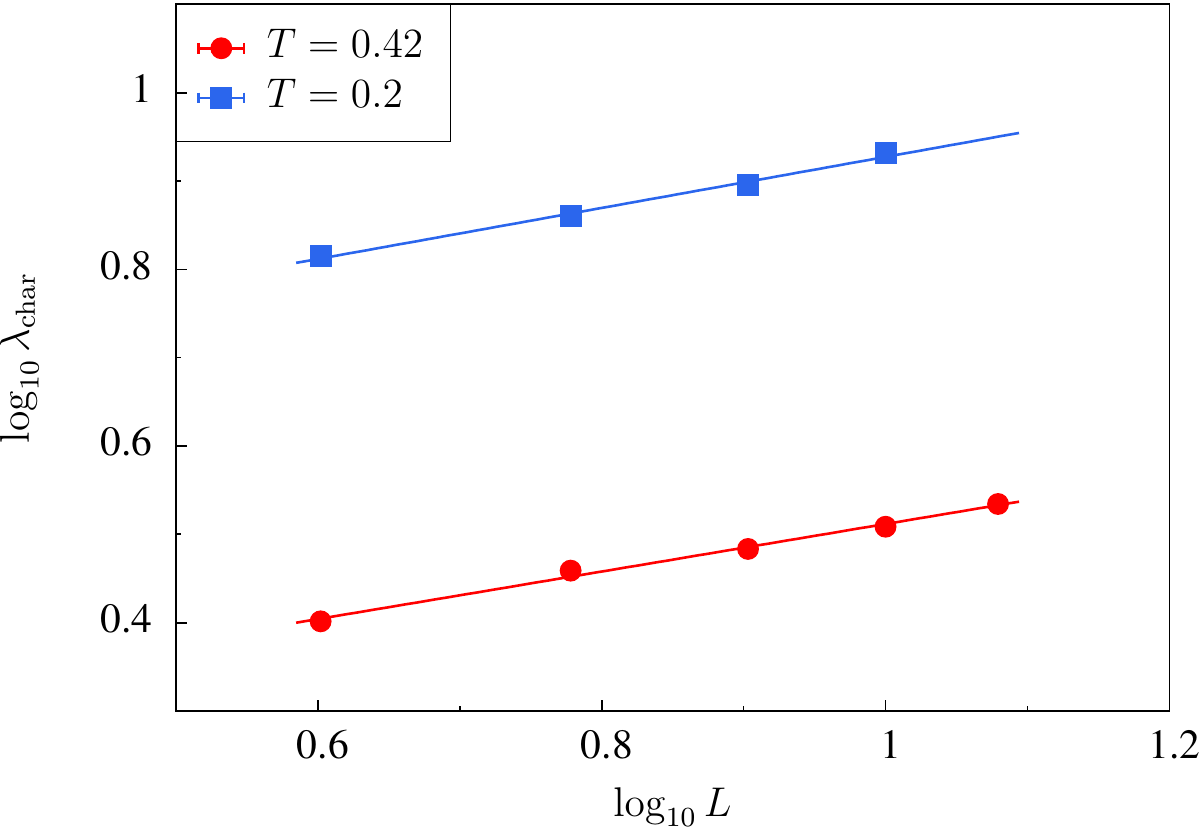}
\caption{
Log-log plot of $\lam_{\rm char}(L)$ vs $L$ for $T=0.2$ and $T=0.42$.
The straight lines represent fits of the form $\lam_{\rm char}(L) \sim a
L^{\theta_{\lam}}$.
}
\label{fig:lamvL}
\end{figure}

Figure \ref{fig:lamvL} shows $\lam_{\rm char}(L)$ vs $\log L$ for
$T=0.2$ and $T=0.42$. Since $\lam_{\rm char}(L)$ is a stiffness measure,
we can extract a new stiffness exponent $\theta_\lam$ from a fit to the
form,
\begin{equation}
\label{eq:thetalam}
\lam_{\rm char}(L) \sim a L^{\theta_\lam} .
\end{equation}  
The values of $\theta_\lam$, given in Table \ref{tab:theta}, are larger
than $\theta$ obtained from the average free energy difference but close
to the value, $0.27$, found in Ref.~\cite{CaBrMo02} using aspect ratio
scaling. Presumably, the asymptotic values of $\theta$ and $\theta_\lam$
are the same. We prefer the larger value, $\theta_\lam$ because it is
obtained from the {\em tail} of the stiffness distribution so we believe
it reflects the large-size behavior more accurately than the average
free energy difference that defines $\theta$. Aspect ratio scaling is
an independent way to minimize finite-size effects and it is
interesting that these two approaches yield the same answer within error
bars.

\begin{figure}[htb]
\begin{center}
\includegraphics[scale=0.68]{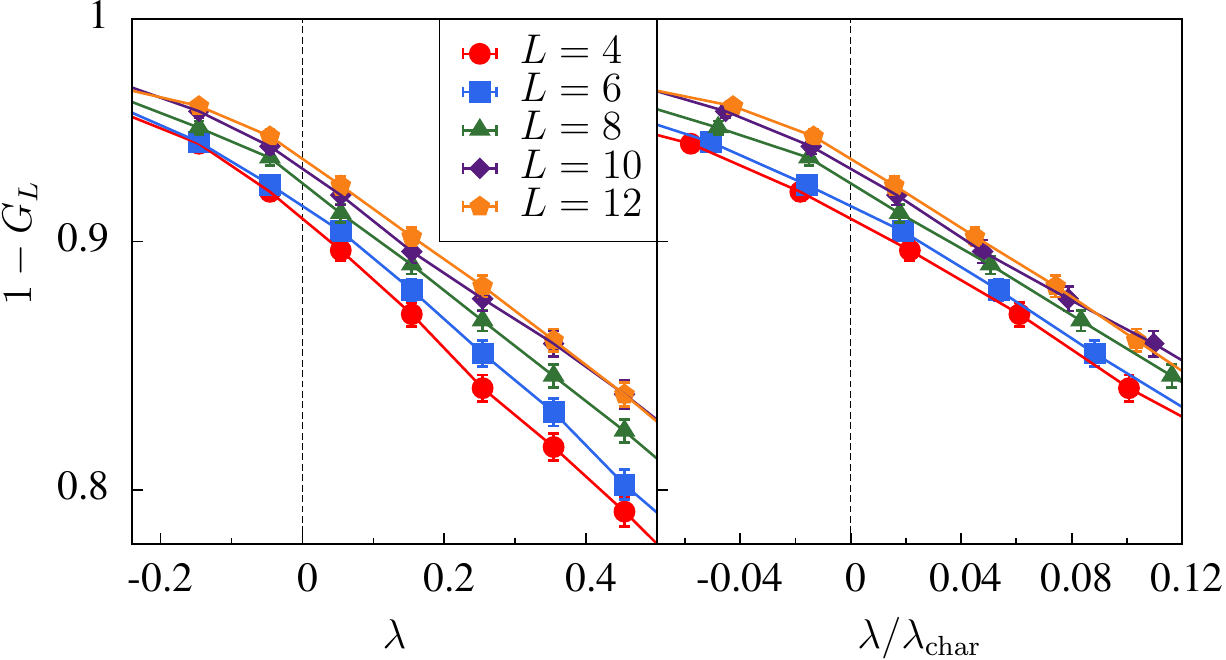}
\caption{
(Left panel) $1-G_L(\lam)$ vs $\lam$ for system sizes $L=4$ through $12$
at $T=0.42$ in the region near $\lam=0$. (Right panel)
$1-G_L(\lam/\lam_{\rm char}(L))$ vs $\lam/\lam_{\rm char}(L)$. Note that
$1-G_L(0)$ increases slowly with $L$.
}

\label{fig:lamccdf42zoom}
\end{center}
\end{figure}

\begin{figure}[htb]
\begin{center}
\includegraphics[scale=0.68]{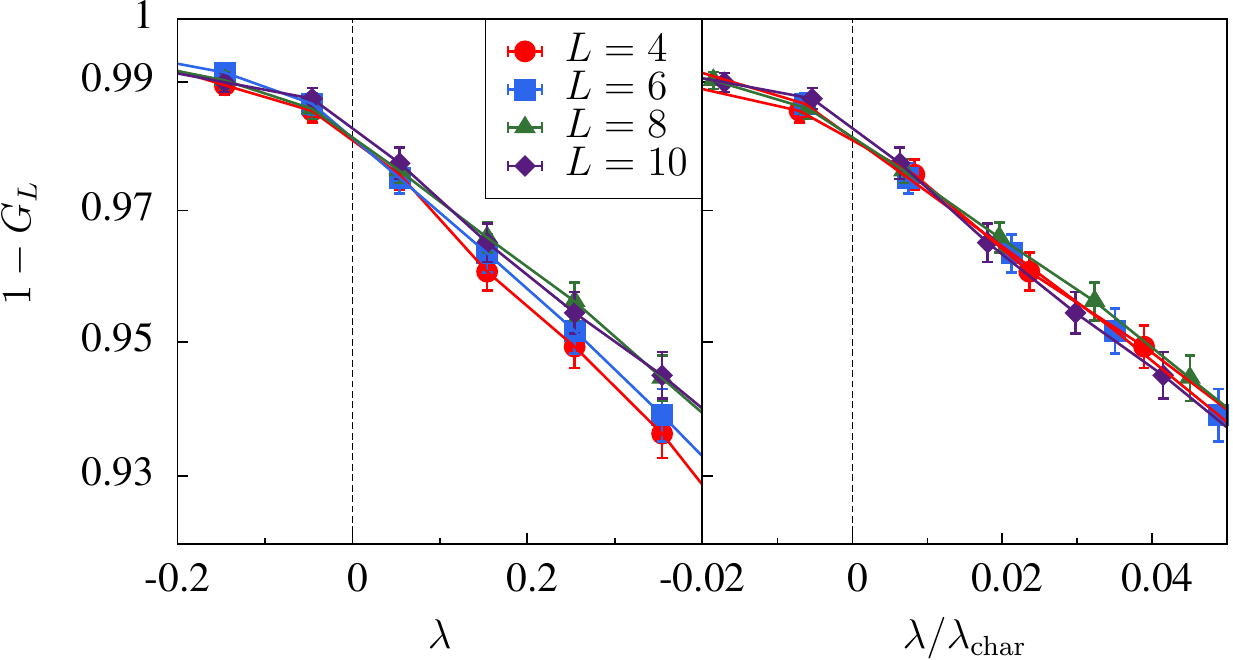}
\caption{
(Left panel)  $1-G_L(\lam)$ vs $\lam$ for system sizes $L=4$ through $10$
at $T=0.2$ in the region near $\lam=0$. (Right panel)
$1-G_L(\lam/\lam_{\rm char}(L))$ vs $\lam/\lam_{\rm char}(L)$.
}
\label{fig:lamccdf20zoom}
\end{center}
\end{figure}

It seems clear that $\lam_{\rm char}(L) \to \infty $ as $L \rightarrow
\infty$. At least on a coarse scale, the full distribution $G_L(\lam)$
also scales with $\lam_{\rm char}(L)$.  A closer look at $G$ near the
head of the distribution for $T=0.42$ shows that the data collapse is
not perfect and there are significant finite-size corrections near
$\lam=0$.  Figure \ref{fig:lamccdf42zoom} shows $1-G_L(\lam/\lam_{\rm
char}(L))$ vs $\lam$ in the region near $\lam=0$ for $T=0.42$. Note that
curves do not collapse perfectly and that $1-G_L(\lam/\lam_{\rm
char}(L))$ appears to be increasing with $L$.  Figure
\ref{fig:lamccdf20zoom} is the same plot for $T=0.2$.  Because  $G_L(0)$
is so small for $T=0.2$, the error bars are too large to discern whether
there is a trend with $L$. A reasonable hypothesis is that there is an
asymptotic $L \rightarrow \infty$ scaling function $G_\infty(z)$ where
$z=\lam/\lam_{\rm char}$ such that $G_L(\lam) \rightarrow
G_\infty(\lam/\lam_{\rm char}(L))$.  The straight line behavior of
$\log(1-G_\infty(z))$  for large $z$ and increasing trend with $L$ for
small $z$  suggests that $G_\infty(0)=0$  and $G_\infty(z)$ is
exponential for $z \gg 1$. In more physical terms, if $G_\infty(z)$
exists and is zero for $z \rightarrow 0^+$, it means that a single
boundary condition dominates the \tbc~ensemble almost surely, i.e., the
dominant boundary condition almost always has a much lower free energy
than the other seven boundary conditions. A more complicated possibility
is that $G_\infty(0^+)>0$.  The consequences of these possibilities for
the RSB vs two-state pictures are discussed in
Sec.~\ref{sec:disc}.

It is noteworthy that for the system sizes accessible to simulations,
$\lam_{\rm char}(L)$ is sufficiently small that the \tbc~ensemble
contains a mixture of several competing boundary conditions for a
substantial fraction of samples.  The disorder average $\po_L$ is
dominated by these samples and is therefore not characteristic of the
large-$L$ behavior when $\lam_{\rm char}(L)$ is expected to be large.
In what follows we circumvent this difficulty by extrapolating first in
$\lam$ and then in $L$, making use of the fact our data contain a
relatively large dynamic range in $\lam$.

\subsection{Order parameter near $q = 0$}
\label{sec:I}

\begin{figure}[htb]
\begin{center}
\includegraphics[scale=0.68]{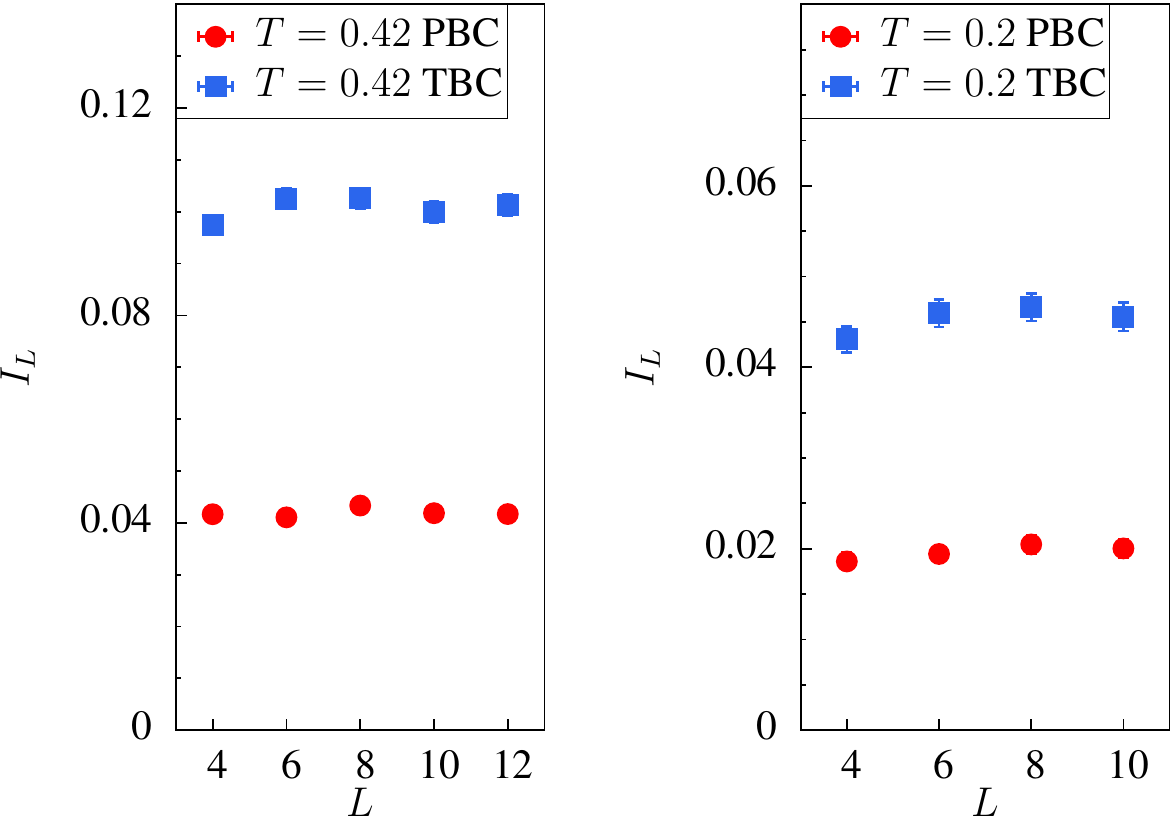}
\caption{
$\po_L$ vs $L$ for PBC and \tbc~at temperature $T=0.42$ (left panel) and
$T=0.2$ (right panel). The data seem independent of system size,
suggesting an RSB interpretation of the data.
}
\label{fig:P0}
\end{center}
\end{figure}

\begin{figure}[htb]
\begin{center}
\includegraphics[scale=0.80]{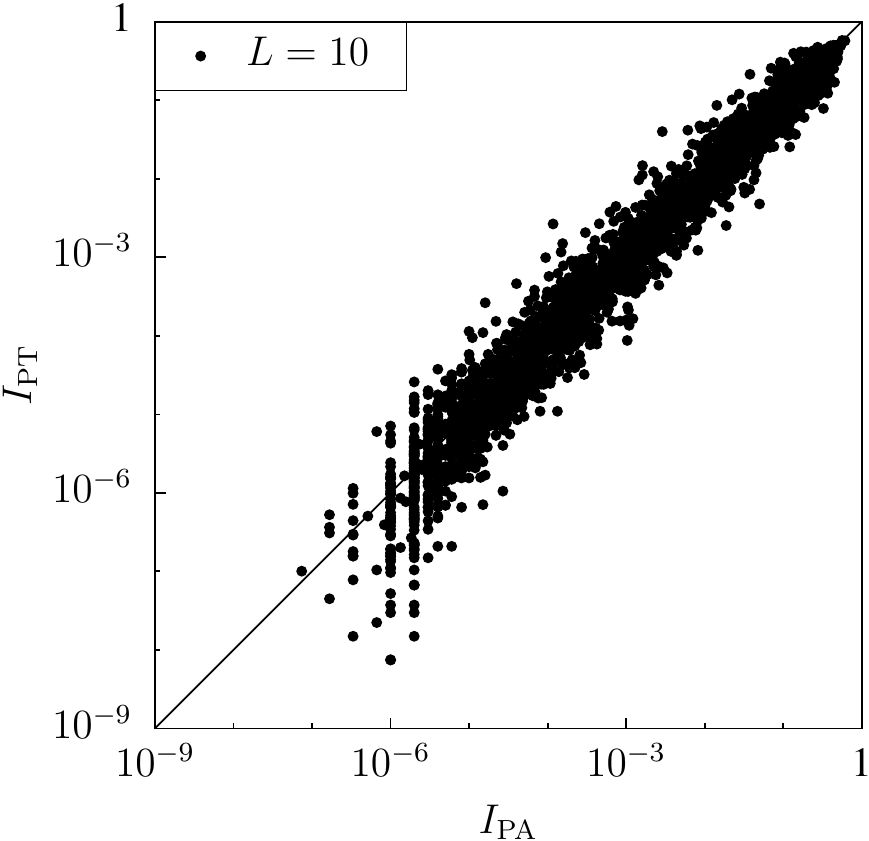}
\caption{
Log-log scatter plot of $\poc$ for PBC for each sample at temperature
$T=0.20$ and $L =
10$ computed with parallel tempering Monte Carlo (PT, vertical axis) and
population annealing Monte Carlo (PA, horizontal line). The data show both
methods yields the same result  for each sample within statistical errors.
}
\label{fig:P010}
\end{center}
\end{figure}

Figure \ref{fig:P0} shows $\po_L$, the disorder average of the
integrated order parameter distribution with $|q|<0.2$, as a function of
size $L$ for temperatures $T=0.2$ and $0.42$, as well as for both PBC
and \tbc.  For both boundary conditions, $\po_L$ is, within error bars,
independent of system size. The results for PBC are identical to those
obtained using parallel tempering Monte Carlo
\cite{katzgraber:01,yucesoy:12}. In fact, in Fig.~\ref{fig:P010} we
show a scatter plot of $\poc$ computed both with population annealing
and parallel tempering Monte Carlo. The data are strongly correlated and  both methods yield  the same results within statistical errors.

The constancy of $\po_L$ has been taken as strong evidence for the RSB
picture because the two-state picture predicts $\po_L$ should decrease as
$L^{-\theta}$. However, in what follows we argue that in
\tbc~ultimately $\po_L \rightarrow 0$ for very large $L$.  On first
glance the results for \tbc~are surprising since $\po_L^{\rm \tbc}$ is
larger by more than a factor of two than $\po_L^{\rm PBC}$.  The
explanation is that for many samples the \tbc~ensemble contains several
boundary conditions with significant weight and the overlap between spin
configurations with different boundary conditions will tend to have
small values of $q$ due to the existence of a relative domain wall.  We
shall return to this important point in Sec.~\ref{sec:disc}.

\subsection{Order parameter near $q = 0$ vs spin stiffness}
\label{sec:Ic}
\begin{figure}[htb]
\begin{center}
\includegraphics[scale=0.65]{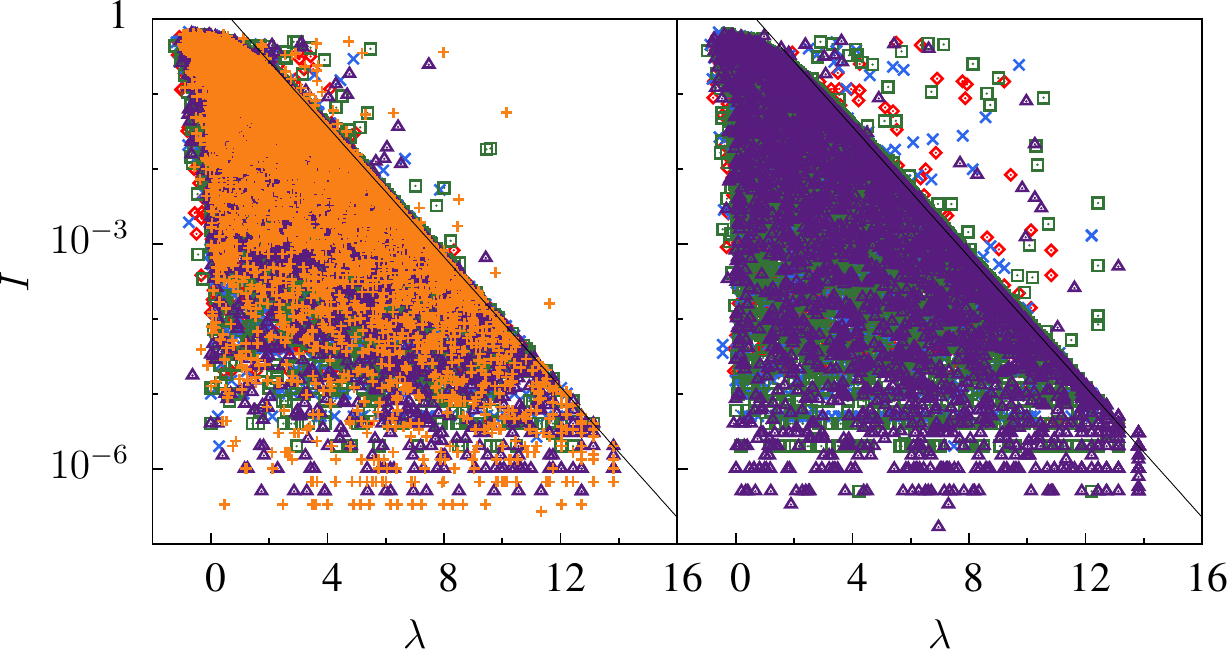}
\caption{(Color online)
Scatter plots showing all disorder realizations for all system sizes at
$T=0.42$ (left panel) and $T=0.2$ (right panel). Each point represents a
sample $\cJ$ located at $x$ coordinate $\lam_{\cJ}$ and $y$ coordinate
$\po_{\cJ}$. Red diamonds represent $L = 4$, blue crosses $L = 6$, green
squares $L = 8$, purple triangles $L = 10$, and orange plus symbols $L = 12$.
}
\label{fig:polam}
\end{center}
\end{figure}

\begin{figure}[htb]
\includegraphics[scale=0.8]{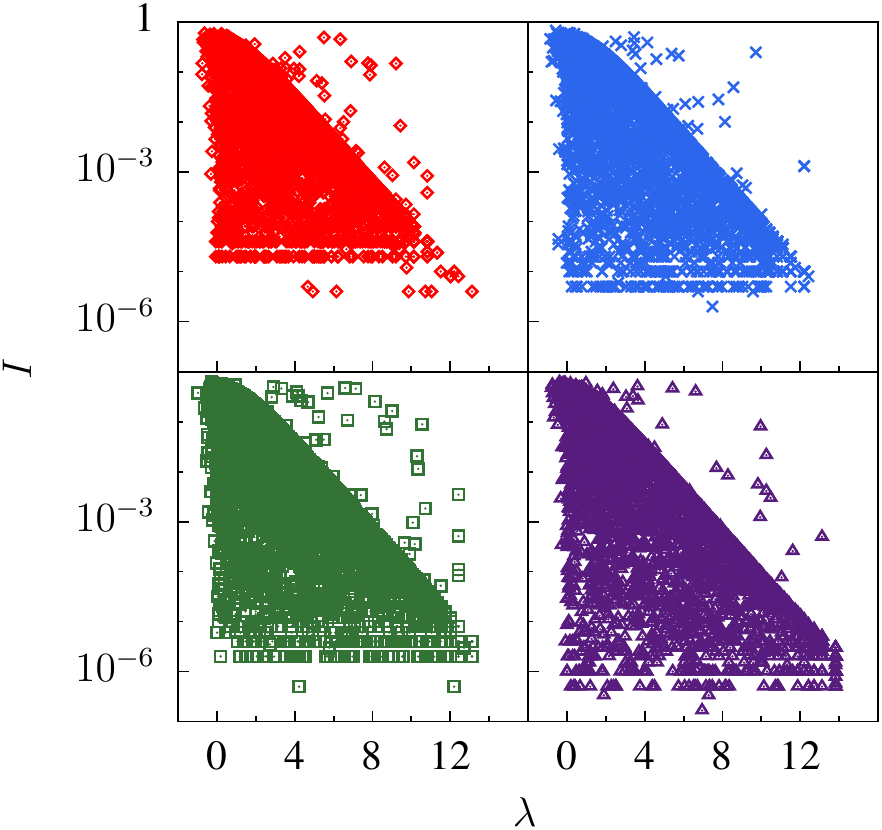}
\caption{(Color online)
Same as Fig.~\ref{fig:polam} but for each system size in a separate
panel and $T=0.2$. Again, red diamonds represent $L = 4$, blue crosses
$L = 6$, green squares $L = 8$, and purple triangles $L = 10$.
}
\label{fig:polamL20}
\end{figure}

Figure \ref{fig:polam} is a scatter plot showing many of the disorder
samples at $T=0.42$ (left panel) and $T=0.2$ (right panel) for all the
sizes studied using \tbc.  Each point on the plot represents a sample
$\cJ$. The $x$ coordinate of the point is
$\lam_{\cJ}=\log\left[f_{\cJ}/(1-f_{\cJ})\right]$ and the $y$ coordinate
is $\po_{\cJ}$. Figure \ref{fig:polamL20} is the same as
Fig.~\ref{fig:polam} but with each system size on a separate plot for
$T=0.2$. The qualitative features of the plots are the same for each
size although, as described above, the distribution of $\lam$'s shifts
to larger values for larger $L$. These figures together with the
behavior of the $\lam$ distribution constitute the main results of our
paper and motivate our conclusion that $\poc \rightarrow 0$ almost
surely as $L \rightarrow \infty$ in thermal boundary conditions.

Samples $\cJ$ for which $\poc$ or $1-f_{\cJ}$ are exactly zero within
the precision of the simulations are not shown on these log-log plots
since $\log \poc=-\infty$ or $\lam_{\cJ}= \infty$. The fractions of such
samples are given in Table \ref{tab:frac}.  Note that actual values
of $\poc$ or $1-f_{\cJ}$ are are never exactly zero for finite systems; zeros correspond to values smaller than can be represented by the finite population sizes used in the simulations.  It is important to note that the trends shown in
Fig.~\ref{fig:polam} continue to hold for the large values of $\lam$
that are omitted from this figure. Including all sizes, there are $216$
samples with $1-f_{\cJ}=0$ for $T=0.42$ and $2996$ such samples for
$T=0.2$.  Of these, only seven samples for $T=0.42$ and $38$ for $T=0.2$
are measured to have nonzero values of $\poc$. The average value of
$\po$ for only those samples with $1-f_{\cJ}=0$ are $2.5 \times 10^{-8}$
and $10^{-4}$ for $T=0.42$ and $T=0.2$, respectively.

\begin{table}
\caption{
Fraction of samples with $\poc=0$ and $f_{\cJ}=1$ for different sizes,
temperatures, and boundary conditions.
\label{tab:frac}
}
\begin{tabular*}{\columnwidth}{@{\extracolsep{\fill}} l l l l l l}
\hline
\hline
\multicolumn{6}{c}{PBC}\\
\hline
\hline
System size $L$ 		  &4 &6 &8 &10 &12  		   \\ \hline
Fraction $\poc=0\ (T=0.42)$ 	  &0.21 &0.19 &0.16 &0.16 &0.19    \\ \hline
Fraction of $\poc=0\ (T=0.2)$ 	  &0.60 &0.57 &0.55 &0.54 &-- 	   \\ \hline
Fraction of $f_{\cJ}=1\ (T=0.42)$ & --&--&-- &-- &-- 		   \\ \hline
Fraction of $f_{\cJ}=1\ (T=0.2)$  & --& --& --& --& --		   \\
\hline
\hline\\
\hline
\hline
\multicolumn{6}{c}{TBC}\\
\hline
\hline
System size $L$                   &4 &6 &8 &10 &12                   \\ \hline
Fraction $\poc=0\ (T=0.42)$       &0.05 &0.04 &0.03 &0.03 &0.03      \\ \hline
Fraction of $\poc=0\ (T=0.2)$     &0.35 &0.33 &0.30 &0.28 &--        \\ \hline
Fraction of $f_{\cJ}=1\ (T=0.42)$ &0.009 &0.009 &0.008 &0.008 &0.010 \\ \hline
Fraction of $f_{\cJ}=1\ (T=0.2)$  &0.15 &0.16 &0.15 &0.15 &--        \\
\hline
\hline
\end{tabular*}
\end{table}

A striking feature of Figs.~\ref{fig:polam} and \ref{fig:polamL20} is
that there is a bounding curve that becomes a straight line for large
$\lam$ with most samples lying {\em below} that curve. Why are there
two classes of samples, with most samples below the curve and a few
above it? We speculate that the samples below the curve have nonzero
values of $\poc$ as a result of the overlap between spin configurations
with {\em different} boundary conditions. The contribution to the
overlap between different boundary conditions in the \tbc~ensemble
cannot exceed $2f(1-f)$ in the limit $f \rightarrow 1$, so that if this
is the primary mechanism producing small overlaps in sample $\cJ$ then
$\log \poc < (-\lam_{\cJ} +\log 2$).  The straight lines in
Fig.~\ref{fig:polam} are defined by $\log \po =( -\lam +\log 2)$.  Thus,
for most samples, we believe that the primary contribution to $\poc$
comes from the overlap between different boundary conditions.  For the
rare samples above the bounding curve, the primary contribution to
$\poc$ must come from small overlaps within the dominant boundary
condition.

\begin{figure}[htb]
\begin{center}
\includegraphics[scale=0.68]{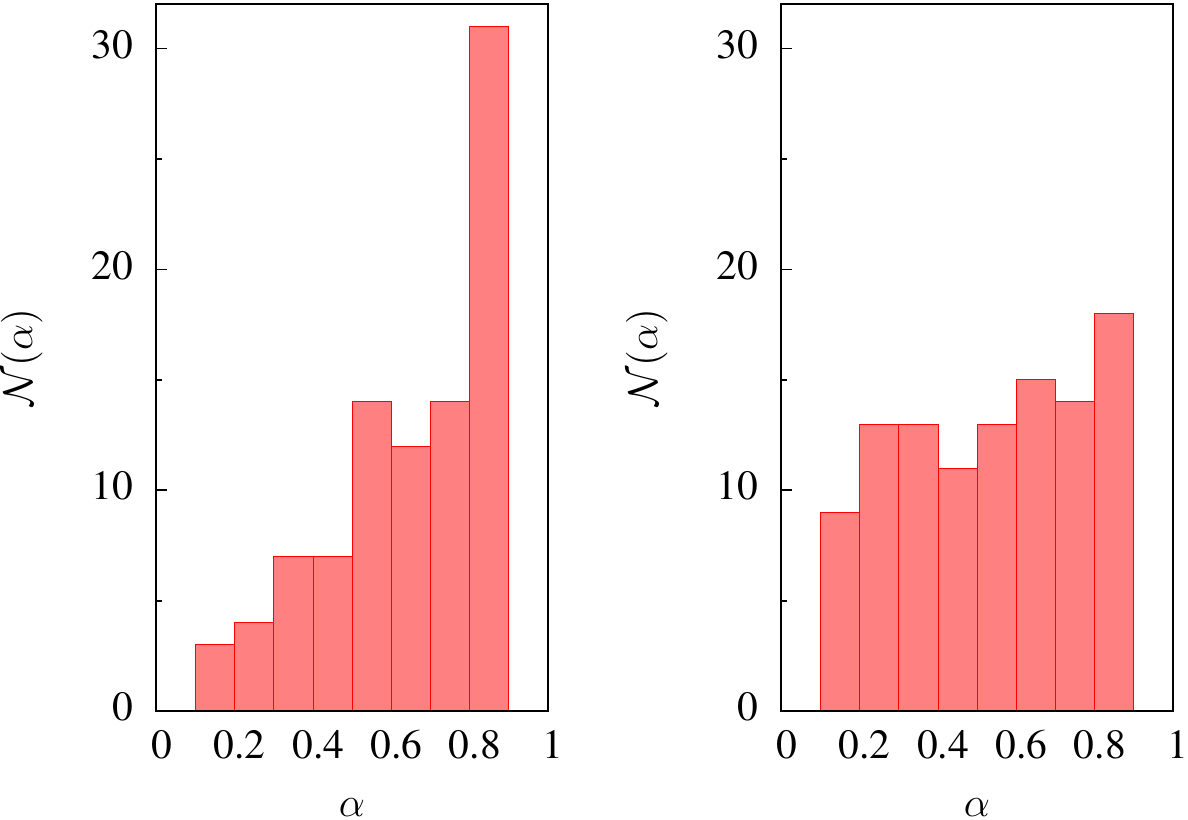}
\caption{
Histogram ${\mathcal N}(\alpha)$ for $\alpha_{\cJ}=\log (\poc
)/[-\lam_{\cJ}+\log(2)]$ for $T=0.42$ (left panel) and $T=0.2$ (right
panel). $\alpha=1$ corresponds to the small values of $\po$ at the
bounding line.
}
\label{fig:logdis}
\end{center}
\end{figure}

A second important feature of Figs.~\ref{fig:polam} and
\ref{fig:polamL20} is that the rare samples above the bounding curve
have $\poc$ roughly uniformly distributed on a logarithmic scale between
the bounding curve and $1$. On a linear scale this means that for
large $\lam$ almost all of these samples have small values  of $\poc$.
Let $\rho(x|y)$ be the conditional probability density for $x=\poc$
conditioned on $y=\lam_{\cJ}$.  If this distribution is exactly log uniform
above the bounding line then the part of the distribution above the line
would take the form,
\begin{equation}
\label{eq:above}
\rho(x|y) = \frac{1-w(y)}{xy} \;\;\;\;\;\;\; 
\mbox{ for } 
\;\;\;\;\;\;\; x>2 \exp(-y),
\end{equation}
where $w(y)$ is the fraction samples below the bounding line. Figure
\ref{fig:logdis} shows histograms of $\poc$ values of the samples of all
sizes that lie above the bounding line for the two temperatures.  The
position $\alpha$ along the horizontal axis is the scaled logarithmic distance
between the bounding line and one. That is,  $\alpha_{\cJ}=-\log
\poc/[\lam-\log (2 )]$ so that  zero corresponds to large values, $\poc
\approx 1$ while one corresponds to $\poc$ on the bounding line.  For
$T=0.2$ the distribution is indeed relatively uniform on a logarithmic
scale while for $T=0.42$ it is skewed to a small value of $\po$.

\section{Discussion}
\label{sec:disc}

Three salient  features of the data are apparent from
Figs.~\ref{fig:lamvL}, \ref{fig:polam}, and  \ref{fig:logdis}.

\begin{enumerate}[I.]

\item{Typical values of the sample stiffness $\lam_{\cJ}$  increase with
system sizes $L$, as described by $\lam_{\rm char}(L)$.}

\item{Most samples have $\poc$ less than a bounding curve described by
$2e^{-\lam}$ for large  $\lam$.}

\item{Samples with $\poc$ above the bounding curve have $\poc$
distributed more or less uniformly in $\log \po$ between the bounding
curve and one.}

\end{enumerate}

We conjecture that these features hold for arbitrarily large $L$ and all
temperatures in the low-temperature phase. Assuming the above statements
are asymptotically correct, we can draw some strong conclusions about
how $\po_L$ behaves for sufficiently large $L$ that $\lam_{\rm char}(L)
\gg 1$.  Note that these system sizes are far larger than are accessible
in our simulations but, given the large dynamic range in $\lam$ we can
extrapolate to these sizes by first extrapolating in $\lam$.  For large
$L$, $\lam_{\cJ}$ is nearly always large according to Statement
(\MakeUppercase{\romannumeral 1}).  Furthermore, due to Statements
(\MakeUppercase{\romannumeral 2}) and (\MakeUppercase{\romannumeral 3}),
$\poc$ is, almost always small when $\lam_{\cJ}$ is large.   Thus $\poc$
is nearly always small when $L$ is large. {\em This conclusion is the
main result of our analysis. It is consistent with two-state pictures
but not consistent with the RSB picture.}

In addition to being consistent with our data, these conjectures are
quite plausible. Statement (\MakeUppercase{\romannumeral 1}) asserts
that $\lam_{\cJ}$ is a measure of sample stiffness and that in the 
low-temperature phase, almost all samples become stiff for large system
sizes.  As discussed above, Statement (\MakeUppercase{\romannumeral 2})
asserts that in \tbc~large values of $\poc$ arise mostly from the
overlap between two different boundary conditions.  Statement
(\MakeUppercase{\romannumeral 3}) asserts that the free-energy cost of a
large excitation in the dominant boundary condition is more or less
uniformly distributed between $\lam/\beta$ and $0$.

We can make the arguments more quantitative using a simple model of how
the disorder average $\po_L$ will behave for large $L$.  Let $\rho( x |
y)$ be the conditional probability density for $\poc=x$ conditioned on
$\lam_{\cJ}=y$.  Based on statements (\MakeUppercase{\romannumeral 2})
and (\MakeUppercase{\romannumeral 3}) we propose the form,
\begin{equation}
\label{eq:rho}
\rho(x|y) = 
w(y) \delta[x,2 \exp(-y)] + \frac{1-w(y)}{x y} \theta[x - 2 \exp(-y)],
\end{equation} 
where $w(y)$ is the fraction of samples at fixed $\lam$ below the
bounding curve,  and $\theta(x)$ and $\delta(x,y)$ are the Heaviside
function and the $\delta$-function, respectively.  The first term
conservatively places all the samples below the bounding curve on the
curve itself. The second term represents the samples above the bounding
curve with the conservative assumption that the distribution of $\poc$
above the curve is uniform on a log scale as in Eq.~\eqref{eq:above}.
For purposes of the following calculation we assume that $w$ is a
constant independent of $y$ but the qualitative conclusions do not
depend on this assumption. Finally, we assume that the distribution of
$\lam$ obeys a size independent form $G_\infty(z)$ for the scaled
variable $z=\lam/\lam_{\rm char}(L)$. Note that we have assumed that the
dependence of $\po_L$ on $L$ is entirely through $\lam_{\rm char}(L)$
and that the conditional probability $\rho(x|y)$ is independent of $L$.
These assumptions yield an explicit expression for $\po_L$ as a function
of $\lam_{\rm char}(L)$,
\begin{equation}
\label{eq:integral}
\po_L = \frac{1}{\lam_{\rm char}(L)} 
\int_0^{\infty} dG_{\infty}(z) \int_0^{\infty} dx~x 
\rho(x|z\lam_{\rm char}(L)).
\end{equation}
Plugging in the ansatz of Eq.~\eqref{eq:rho} and an exponential form for
the scaled $\lam$ distribution, $1-G_\infty(z) = e^{-z}$, yields a
somewhat complex expression involving exponential integrals whose
asymptotic large $\lam$ behavior is,
\begin{equation}
\label{eq:po}
\po_L \sim 
\frac{1}{ \lam_{\rm char}(L)} 
\left[2w  + (1-w) \log( \lam_{\rm char}(L)) \right].
\end{equation} 
Using Eq.~\eqref{eq:thetalam} and assuming that asymptotically
$\theta_\lam=\theta$, we recover the prediction of the two-state picture
that $\po_L \sim L^{-\theta}$, however, with a logarithmic correction that
arises from the assumption of a log-uniform distribution for
$\po_{\cJ}$.

While the above assumptions lead to an explicit asymptotic expression
for $\po_L$ as a function $L$, this expression should not be taken too
seriously. However, the qualitative conclusion  that $\poc \rightarrow
0$ for almost all samples as $L \rightarrow \infty$ is robust and
depends only on the asymptotic validity of statements
(\MakeUppercase{\romannumeral 1}) --- (\MakeUppercase{\romannumeral 3})
above.

Given that $\lam_{\rm char}(L)$ is increasing with $L$ and that $\po$
decreases with increasing $\lam$, why is $\po_L$ nearly constant for the
sizes studied in our \tbc~simulations of the 3D EA model?  We believe
this conundrum can be explained, at least in part, by  the fact that the
main contribution to $\po_L$ comes from samples with small values of
$\lam$. For $T=0.2$, more than half the contribution to $\po_L$ comes
from samples with $\lam < 1$ and more than $80\%$ from $\lam<2$, and
these fractions are even higher for $T=0.42$. The head of the $\lam$
distribution, has very little dependence on $L$, as can be seen in
Figs.~\ref{fig:lamccdf42} and \ref{fig:lamccdf20}. Furthermore, the
bounding curves in Fig.~\ref{fig:polam} are nearly flat in the small
$\lam$ region. Thus several effects come into play in keeping $\po_L$
nearly independent of $L$. First, the main contribution to $\po_L$ is
from samples with small stiffness.  Second, the fraction of samples with
small stiffness decreases by only a small amount for the sizes studied
and, finally, $\po$ does not depend much on $\lam$ for small $\lam$.
One would have to go to much larger sizes before $\po_L$ would decrease
according to the predicted asymptotic power law $L^{-\theta}$.

In the foregoing, we have assumed that $G_L(0) \rightarrow 0$ as $L
\rightarrow \infty$ or, equivalently, if $G_\infty(z)$ exists,
$G_\infty(0^+)=0$.  We now consider the consequences of an alternate
assumption that $\lam_{\rm char}(L) \rightarrow \infty$ and
$G_\infty(z)$ exists but $G_\infty(0^+)>0$. This possibility cannot be
ruled out by the data although if it holds, it appears that
$G_\infty(0^+)$ is quite small. In physical terms $G_\infty(0^+)>0$
means that even for very large sizes, a fraction $G_\infty(0^+)$ of
samples has a mixed ensemble of boundary conditions in \tbc~while the
remaining samples have only a single boundary condition in the
\tbc~ensemble.  This scenario would imply that the 3D EA model in
\tbc~is  divided into two classes of disorder realizations, one of
which, with weight $(1-G_\infty(0^+))$, has $\po_L=0$ and  the other,
with weight $G_\infty(0^+)$, has $\po_L>0$.  This possibility seems
unlikely but is not contradicted by the data. It has no straightforward
explanation in either two-state or RSB pictures.

Our hypothesis is that thermal boundary conditions and periodic boundary
conditions have the same behavior in the limit of large system sizes.
We use thermal boundary conditions as a tool to improve the
extrapolation to large system sizes from the very small system sizes
accessible in simulations.  It is known that coupling dependent boundary
conditions are not equivalent to periodic boundary conditions and are
not suitable for understanding properties of the spin glass phase
because they could be used to select a single pure state even if
coupling independent boundary conditions  admit many pure states.  The
status of thermal boundary conditions with regard to coupling dependence
is not clear. On the one hand, the \tbc~ensemble contains different
mixtures of boundary conditions for different choices of couplings.  On
the other hand, the particular mixture of the 8 boundary conditions
is chosen by the system itself and is not imposed externally.  As
discussed in Sec.~\ref{sec:tbc}, our intuition is that \tbc~minimizes
finite size effects rather than introducing spurious physics but this
question requires further investigation.  In any case, we have provided
compelling evidence that the 3D EA model in thermal boundary conditions
is best described by a picture with a single pair of pure states in each
finite volume.

\section{Summary}
\label{sec:summary}

We have introduced two techniques with the  aim of extrapolating to
the large system-size behavior of finite-dimensional spin glasses at low
temperature.  First,  we use thermal boundary conditions to minimize the
effect of domain walls induced by boundary conditions.  Second, we use a
natural measure of sample stiffness defined within thermal boundary
conditions and extrapolate to large values of the sample stiffness.  By
noting that the sample stiffness increases with system size we then
obtain an extrapolation in system size.  The dynamic range in sample
stiffness in the data is sufficiently large that a qualitative
extrapolation is readily apparent.    The  conclusion is that nearly all
large samples will have essentially no weight in the overlap
distribution near zero overlap.  The analysis also explains why this
qualitative behavior cannot be seen  using a direct extrapolation in
system size for the small sizes studied.  Our conclusions are consistent
with  two-state pictures but are inconsistent with the mean field,
replica symmetry breaking picture.

Our results hold for thermal boundary conditions.  We believe that
thermal boundary conditions are equivalent to other coupling independent
boundary conditions so that our  conclusions about the infinite volume
limit also apply to the more familiar periodic boundary conditions.
However, it is important to investigate the equivalence of thermal and
periodic boundary conditions.

Our numerical simulations used population annealing.  This Monte Carlo
algorithm has not been used before for large-scale studies in
statistical physics.  We found that it is an effective computational
tool with several advantages over  parallel tempering, the standard
computational method in the field. We believe that population annealing
will be useful for other hard problems in statistical physics and
related fields.  Similarly, thermal boundary conditions and
extrapolating in sample stiffness are general methods that should be
useful in studying other finite-dimensional disordered systems.

\begin{acknowledgments}

H.G.K.~acknowledges support from the NSF (Grant No.~DMR-1151387) and
would like to thank Georg Schneider \& Sohn for creating TAP4.  J.M.~and
W.W.~acknowledge support from NSF (Grant No.~DMR-1208046). We are
grateful to Alan Middleton, Dan Stein, and Chuck Newman for reading an
early version of the manuscript and for useful discussions.  We
acknowledge the contribution of Burcu Yucesoy in providing comparison
data from parallel tempering simulations and for useful discussions. We
thank the Texas Advanced Computing Center (TACC) at The University of
Texas at Austin for providing HPC resources (Stampede Cluster), ETH
Zurich for CPU time on the Brutus and Euler clusters, and Texas A\&M
University for access to their Eos and Lonestar clusters. We especially
thank O.~Byrde for beta access to the Euler cluster.

\end{acknowledgments}

\bibliographystyle{apsrevtitle}
\bibliography{references,refs}

\end{document}